\documentclass[aip,jcp,reprint,showpacs,floatfix,twocolumn,superscriptaddress]{revtex4-1}

\pdfoutput=1

\usepackage{natbib} 
\usepackage{amsfonts} 
\usepackage{amsmath} 
\usepackage{amssymb} 
\usepackage{graphicx}

\makeatletter
\newcommand*{\balancecolsandclearpage}{%
  \close@column@grid
  \clearpage
  \onecolumngrid
}

\begin{document}

\title{Diffusion- and reaction-limited cluster aggregation revisited}

\author{Swetlana Jungblut}
\email{swetlana.jungblut@tu-dresden.de}
\affiliation{Physikalische Chemie, TU Dresden, Bergstra{\ss}e 66b, 01069 Dresden, Germany}
\author{Jan-Ole Joswig}
\affiliation{Theoretische Chemie, TU Dresden, Bergstra{\ss}e 66c, 01069 Dresden, Germany}
\author{Alexander Eychm{\"u}ller}
\affiliation{Physikalische Chemie, TU Dresden, Bergstra{\ss}e 66b, 01069 Dresden, Germany}

\date{\today}

\begin{abstract}
  
We simulated irreversible aggregation of non-interacting particles and of particles interacting via repulsive and attractive potentials explicitly implementing the rotational diffusion of aggregating clusters. Our study confirms that the attraction between particles influences neither the aggregation mechanism nor the structure of the aggregates, which are identical to those of non-interacting particles. In contrast, repulsive particles form more compact aggregates and their fractal dimension and aggregation times increase with the decrease of the temperature. A comparison of the fractal dimensions obtained for non-rotating clusters of non-interacting particles and for rotating clusters of repulsive particles provides an explanation for the conformity of the respective values obtained earlier in the well established model of diffusion-limited cluster aggregation neglecting rotational diffusion and in experiments on colloidal particles.

\end{abstract}
 
\maketitle

\section{Introduction \label{intro}}

The process of the formation of highly porous low-density non-equilibrium structures by diffusion- and reaction-limited cluster aggregation (DLCA \cite{kolb:1983,meakin:1983a, meakin:1984} and RLCA \cite{jullien:1984,family:1985,brown:1985,meakin:1987,meakin:1988}, respectively) has been extensively studied in the 1980's, when the concepts of the model, supported by the experimental results \cite{weitz:1984a, weitz:1984b}, were introduced. In particular, it was proposed to consider the aggregation of nanoparticles into fractal disordered structures as a process, in which particles and clusters of particles diffuse through the surrounding medium and stick together irreversibly either on every collision (DLCA) or with a certain probability (RLCA). Experimentally, the model was first implemented with gold colloidal particles sterically stabilized by electrostatic repulsion whose aggregation was induced by the reduction of the interparticle repulsion through the neutralization of the involved particles \cite{weitz:1984a, weitz:1984b}. Later on, the universality of the findings was demonstrated on the examples of the aggregation of polystyrene and silica particles \cite{lin:1989,lin:1990a,lin:1990b}. One of the central conclusions of these investigations was that, considering repulsive and non-interacting particles, there is a universal limit of the porosity of the structures which is given by the aggregation of non-interacting particles. The characteristic parameter, to which this limit applies, is the fractal (Hausdorff) dimension. Its value indicates how effectively the structure in question fills the available space. In three dimensions, the values of the fractal dimensions of structures formed by cluster aggregation were found to vary from $d_f^{\rm DLCA}=1.7-1.8$ to $d_f^{\rm RLCA}=1.9-2.1$. At high dilutions, the fractal structure of the aggregates is independent of the initial concentration of the aggregating particles, but starts to increase when the density becomes sufficient to form percolating networks with a fractal dimension of $d_f^{\rm perc}=2.5$.

In principle, the model provides a reasonable scenario of nanoparticle aggregation emerging, for instance, as one of the steps in the aerogel production \cite{mohanan:2005,bigall:2009,liu:2015exp,wen:2016,niederberger:2017,rechberger:2017,ziegler:2017,cai:2018}, in which the destabilization of nanoparticles suspended in a solution induces their aggregation into disordered networks. The recent advances in nanotechnology and the expansion of the research associated with the topic of nanoparticle aggregation revealed, however, some deficiencies of the model. For instance, an experimental study \cite{chakrabarty:2009,sanderchakrabarty:2010,chakrabartyReply:2010} observed porous structures of aggregated carbon particles, which were less compact than those realized within the DLCA, but could not provide an explanation for this observation. Recently, we studied \cite{jungblut:2019} the process of DLCA taking rotational diffusion of the aggregates explicitly into account (rDLCA) and found that the fractal dimension of rotating aggregates is lower than the one obtained within the standard DLCA. Previously, the significance of the rotational diffusion was recognized only by the analysis of the experimental results \cite{lindsay:1988,lindsay:1989} but not in the determination of the cluster structure. The reason for this neglecting provided one of the earlier computer simulation studies \cite{meakin:1988}, which claimed that the rotational diffusion of the clusters, implemented implicitly by a random selection of the relative orientations of aggregating clusters, does not influence the fractal dimension of the forming structures. Hence, subsequent investigations of DLCA \cite{meakin:1985,hasmy:1995b,lattuada:2003a, lattuada:2003, rottereau:2004a, rottereau:2004b,diezOrrite:2005,babu:2008,heinson:2010} restricted the motion of the clusters to its translational component. In this work, we demonstrate that the structures found in experiments supporting the predictions of classical DLCA \cite{weitz:1984a, weitz:1984b} can be interpreted as a result of rDLCA with residual interparticle repulsion. Furthermore, our study reveals that the aggregation of non-interacting and attractive particles proceeds similarly and results in identical structures. Considering rather dilute systems, we find that the fractal dimension of non-percolating clusters is independent of the volume fraction of particles initially present in the solution. We also confirm that the variation of the temperature affects the structure of the aggregates only if they are formed by repulsive particles. In this case, corresponding to the RLCA, the decrease of the temperature reduces the likelihood that two repulsive particles approach each other close enough to form a bond and hence the fractal dimension of the aggregating clusters increases.                    

The article is organized as follows. We start with the presentation of the details of the performed simulations and 
then discuss the results. In the discussion, we concentrate on the fractal and local structures of the aggregates formed by attractive, non-interacting, and repulsive particles at three different temperatures. Since, as expected, the structure of the aggregating clusters is essentially independent of the initial particle density, we present the analysis of the structures for one data set in the main text and provide the other data as supplementary material. We finish the article with a 
summary of our findings.

\section{Simulation details \label{simdetails}}

We simulated the aggregation of attractive, non-interacting, and repulsive particles. The choice of the functional form of the interparticle potential was inspired by the well-known repulsive Yukawa potential between two likely charged particles and reads   
\begin{equation}
u(r)=C_1 \frac{\exp(-C_2 r)}{r}, \qquad {\rm if} \  r>1, 
\end{equation}
where $C_1$ and $C_2$ are constants set to $\{\pm1, 0\}$ and $2$, respectively. Both the attractive and repulsive interparticle interactions were truncated at $r_t=4$ and shifted by $u(r_t)$. Considering repulsive interactions, we identify the constant $C_2$ as the Debye-H{\"u}ckel length 
\begin{equation}
\kappa=\sqrt{8 \pi \lambda_{\rm B} N_{\rm A} I}, 
\end{equation}
where $\lambda_{\rm B}$ is the Bjerrum length of the surrounding medium, $N_{\rm A}$ is the Avogadro number, and $I$ is the ionic strength of the solution. The magnitude of the repulsion is related to the Debye-H{\"u}ckel and Bjerrum lengths via 
\begin{equation}
C_1=\frac{q^2 \lambda_{\rm B} \exp(\kappa \sigma)}{(1+\kappa \sigma/2)^2},
\end{equation}
where $q$ is the charge of a particle.

Throughout the paper, all distances are given in units of the particle diameter $\sigma$ and the time in $\tau = \sigma \sqrt{m/k_{\rm B}T}$. The energy scale is set in terms of the thermal energy $k_{\rm B}T$  with $k_{\rm B}$ being the Boltzmann constant and $T$ the temperature.  
We studied the evolution of the system in a canonical $NVT$ ensemble with temperature controlled by the Langevin thermostat for rigid body dynamics \cite{davidchack:2015}, which couples on the translational as well as rotational degrees of freedom. The mass of a single particle $m$ was set to unity, and hence the mass of an aggregate was equal to $n$, the number of particles it contained. The friction coefficients of the Langevin thermostats are related to the single-particle translational and rotational diffusion constants, set to $D_t = 0.1$ and $D_r=0.5D_t$, via $\gamma_{t|r}=k_{\rm B}T/D_{t|r}$. The evolution of the system is integrated with a time step of $\Delta t = 0.001$. We considered three different temperatures $T=\{0.8, 1, 1.2\}$. 
The particles were confined to a cubic box with the edge $L=60$ and periodic boundary conditions were applied in all directions. The number of particles $N$ varied between $640$ and $2800$, yielding a set of volume fractions $\varphi_i$=$\{0.0015514$, $0.00193925$, $0.00232711$, $0.00290888$, $0.00387851$, $0.00484814$, $0.00581776$, $0.00678739\}$.    
Simulating non-reversible aggregation, we assume that particles and clusters collide inelastically at the given cutoff distance, which was set to $r_c=1$, and continue their movement as rigid bodies with the translational and angular momenta conserved. The rotations of the aggregates were implemented using quaternions \cite{miller:2002}.
For each density, we simulate the aggregation of $100$ realizations of initially disordered systems obtained from equilibrated simulations of strongly repulsive ($C_1=-10$) particles. Initial particle velocities were chosen from the Maxwell-Boltzmann distribution. Each run continues until all particles connect into a single aggregate. Along the runs, we monitor the size and the shape of the clusters containing more than one particle.

\section{Results and discussion  \label{results}}

\begin{figure*}[bt]
\begin{center}
\includegraphics[clip=,width=1.99\columnwidth]{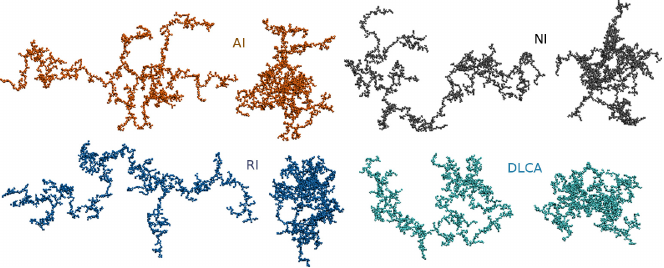}
\caption{\label{fig:snapshot} Snapshots \cite{vmd:1996} of representative clusters formed from the same initial configuration by the attractive (AI) and repulsive (RI) particles as well as by non-interacting particles with (NI) and without (DLCA) rotational diffusion. Each aggregate is presented as two projections onto the planes parallel and perpendicular to the main principal axis.} 
\end{center}
\end{figure*}
In our analysis, we consider the aggregates formed by the attractive, non-interacting, and repulsive particles at three different temperatures and emphasize the impact of rotational diffusion by comparing the structures of the clusters of non-interacting particles aggregated with and without rotational diffusion. Representative snapshots of the final aggregates evolved from the same initial configuration at different conditions, shown in Fig.~\ref{fig:snapshot}, illustrate mainly the difference between the aggregates formed with and without rotational diffusion. A detailed examination of the structures, presented in the following, reveals, however, that only the attractive and non-interacting particles aggregate into similar structures. The clusters formed by the repulsive particles, although visibly similar to the aggregates formed by the attractive and non-interacting particles, have a fractal dimension, which is closer to the one of non-rotating aggregates. The local arrangement of repulsive particles in a cluster resembles, however, neither the distribution of non-interacting particles in a rotating cluster nor the structures evolved in non-rotating aggregates.  

\subsection{Fractal dimension}

\begin{figure}[tb]
\begin{center}
\includegraphics[clip=,width=0.99\columnwidth]{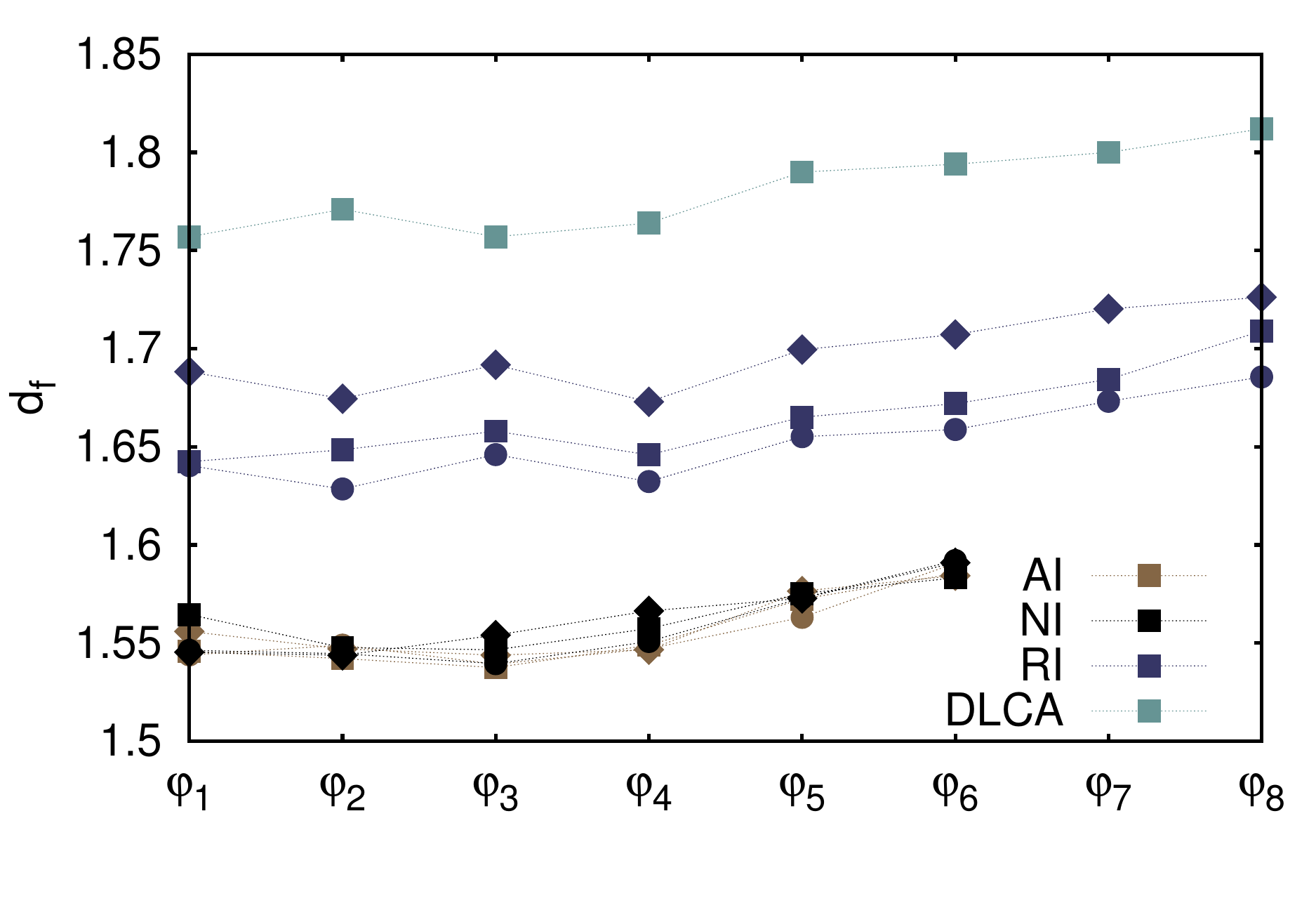}
\caption{\label{fig:df_all} Fractal dimensions of aggregates formed by attractive (AI), repulsive (RI), and non-interacting (NI) particles at $T=0.8$ (diamonds), $1.0$ (squares), $1.2$ (circles) for varying initial particle volume fractions $\varphi_i$. The data for DLCA without rotational diffusion are presented for $T=1.0$ only. The values of fractal dimension are obtained from fitting Eq.~(\ref{rofgscaling}) to the data in the range restricted to $5<n<600$. Uncertainty of all fits is of the order $\pm 0.001$. Lines connecting the data points are guides to the eyes.} 
\end{center}
\end{figure}

One of the standard methods to determine the fractal dimension of porous materials relates the mass of the aggregates to their radius of gyration,   
\begin{equation}
R_g = \sqrt{\lambda_1^2 + \lambda_2^2 + \lambda_3^2},
\end{equation} 
which we compute from the eigenvalues $\lambda_i$ of the gyration tensor extracted from the positions of all particles in the aggregate: 
\begin{equation}
G =\frac{1}{n} \sum_{i=1}^n\left ( \begin{matrix}
  x_ix_i & x_iy_i & x_iz_i \\
  y_ix_i & y_iy_i & y_iz_i \\
  z_ix_i & z_iy_i & z_iz_i
 \end{matrix} \right ).
\end{equation}
The mass of a cluster with fractal structure, which, in our case, is equal to the number of particles it contains, depends on its radius of gyration via  
\begin{equation}
n=kR_g^{d_f}, \label{rofgscaling}
\end{equation} 
where $k$ is a geometric factor, used along with $d_f$ to fit the relation to the data. In order to obtain an 
accurate estimation of the fractal dimension, we collected the data for all cluster sizes appearing in the system in 
the course of the aggregation process. Then, we fitted the data in the size range $5 < n < 600$, which ensures that an increase of 
the fractal dimension of the aggregates due to their overlaps through periodic boundary conditions is not taken into account. 
Figure~\ref{fig:df_all} summarizes the values of the obtained fits. As expected, the data confirms that the fractal 
structure of the aggregates formed by non-interacting particles depends neither on the volume fraction of the particles 
initially present in the system nor on the temperature of the system. Evidently, the same conclusion can be drawn 
for the attractive particles, which aggregate into fractal 
structures indistinguishable from those formed by non-interacting particles. In contrast, repulsive particles arrange 
themselves into structures with higher fractal dimensions. Confirming the RLCA scenario, we observe that the fractal aggregates become less compact with 
increasing temperature, which is consistent with the fact that both the kinetic energy of particles aggregating against a repulsive force used to counteract this force and hence the probability to form a bond between the particles increase.    
Figure~\ref{fig:df_all} further indicates that, as the temperature of the system decreases, the fractal dimension of clusters formed by repulsive particles with rotational diffusion approaches the value of $d_f=1.7-1.8$, which is the fractal dimension of aggregates formed via the conventional DLCA. The latter was shown to agree well with the fractal dimension of experimentally prepared aggregates \cite{weitz:1984a, weitz:1984b}. This agreement supported the conclusion, drawn by an earlier study \cite{meakin:1988} mentioned in the introduction, that the rotational diffusion of clusters does not influence the aggregates' structure. 
Aside from that, fractal clusters observed in experiments diffused both translationally and rotationally. On the basis of our findings, we hypothesize that the gold colloids prepared as non-interacting particles experienced spurious repulsion, which lead 
to a coincidental match of the fractal dimensions of the rotating aggregates formed by repulsive particles and 
of the non-rotating aggregates constructed by non-interacting particles.     

\begin{figure}[tb]
\begin{center}
\includegraphics[clip=,width=0.99\columnwidth]{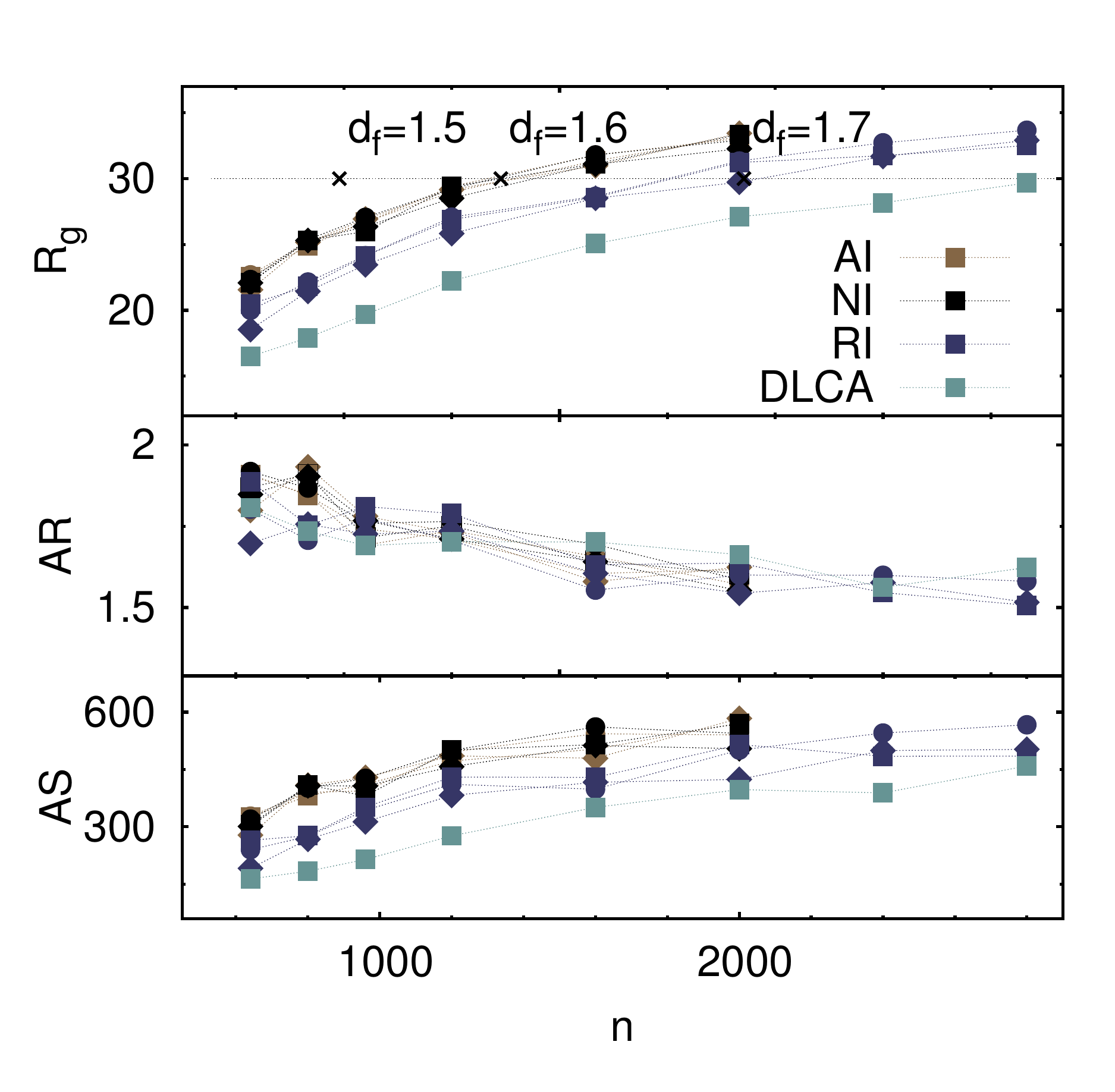}
\caption{\label{fig:rofg_finalaggregates} Radius of gyration (top), aspect ratio (middle), and aspherity parameter (bottom) of the final aggregates formed by attractive (AI), repulsive (RI), and non-interacting (NI) particles at $T=0.8$ (diamonds), $1.0$ (squares), $1.2$ (circles). The data for DLCA without rotational diffusion are presented for $T=1.0$ only. Crosses on the horizontal gray line at $L/2$ indicate the sizes of percolating clusters estimated from Eq.~(\ref{ncofdf}) with different fractal dimensions as labeled. Lines connecting the data points are guides to the eyes.} 
\end{center}
\end{figure}
The variation of the radii of gyration of final aggregates with the number of particles they contain is plotted in Fig.~\ref{fig:rofg_finalaggregates} for different interparticle interactions. For each initial density considered, a higher fractal dimension, observed for clusters of repulsive particles and non-rotating aggregates, is associated with the aggregates that are more compact than those formed by the attractive and non-interacting particles. There are no clear indications to the temperature induced effects on the cluster size, although, from the evaluation of the fractal dimensions, we would expect such effects for the aggregates formed by repulsive particles. 
Otherwise, the increase of the size of final aggregates with the number of particles they contain is in agreement with the consideration on the percolation transition in fractal clusters. Although an appearance of a percolating cluster in our system is highly improbable due to the formation of rigid bonds, we expect to approach the percolation transition when the cluster size becomes comparable with the simulation box size as it is the case for the final aggregates presented in Fig.~\ref{fig:rofg_finalaggregates}. These values further confirm our estimates of the fractal dimensions of the clusters. Crosses on the horizontal line indicating the dimensions of the simulation box stand for the estimated size of a percolating aggregate with a certain fractal dimension. The respective number of particles needed to be initially present in a system of size $L^3$ to observe aggregation into a percolating structure, 
\begin{equation}
n^c=\frac{6}{\pi}L^{d_f}, \label{ncofdf}
\end{equation}
follows from the assumption that $R_g^c=(\sigma/2)\varphi^{1/(d_f-3)}$, which is the size of percolating clusters for given initial volume fraction derived in previous investigations of DLCA \cite{carpineti:1992, bibette:1992a}, is equal to $L/2$. 
Evidently, final aggregates considered in Fig.~\ref{fig:rofg_finalaggregates} approach the size of the simulation box in a range of fractal dimension 
values, which is also obtained from the fit of the aggregating cluster sizes to Eq.~(\ref{rofgscaling}).

The eigenvalues of the gyration tensor can further be used to define the aspherity parameter \cite{theodorou:1985} of the aggregates, 
\begin{equation}
AS = \lambda_1^2-0.5(\lambda_2^2 + \lambda_3^2), 
\end{equation}
which approaches zero for a perfectly spherical object or a spherically symmetric distribution of particles in a cluster. Additionally, a mapping of the gyration tensor on the inertia tensor \cite{vymetal:2011} provides the radii of gyration around the principal axes of an aggregate. Averaging over the radii of gyration corresponding to the two smallest eigenvalues of the inertia tensor, we define the aspect ratio of an aggregating cluster as the ratio of this averaged value to the radius around the axis corresponding to the largest eigenvalue of the inertia tensor: 
\begin{equation}
AR = \sqrt{\frac{\lambda_1^2 + 0.5(\lambda_2^2 + \lambda_3^2)}{\lambda_2^2 + \lambda_3^2}}. 
\end{equation}
The combinations of aspect ratios and aspherity parameters presented in Fig.~\ref{fig:rofg_finalaggregates} indicate that the distribution of particles inside the clusters becomes less spherically symmetric with increasing cluster size, while the elongation of the aggregates decreases. Different to the observation of a previous study \cite{fry:2004}, we find that the aspect ratio of clusters is independent of the kind of interparticle interaction. Such dependence is visible, however, in the distribution of particles inside a cluster, which becomes comparably more spherically symmetric, when individual particles repel each other or the aggregation of non-interacting clusters proceeds without rotational diffusion.

\subsection{Local structure}
\begin{figure}[tb]
\begin{center}
\includegraphics[clip=,width=0.99\columnwidth]{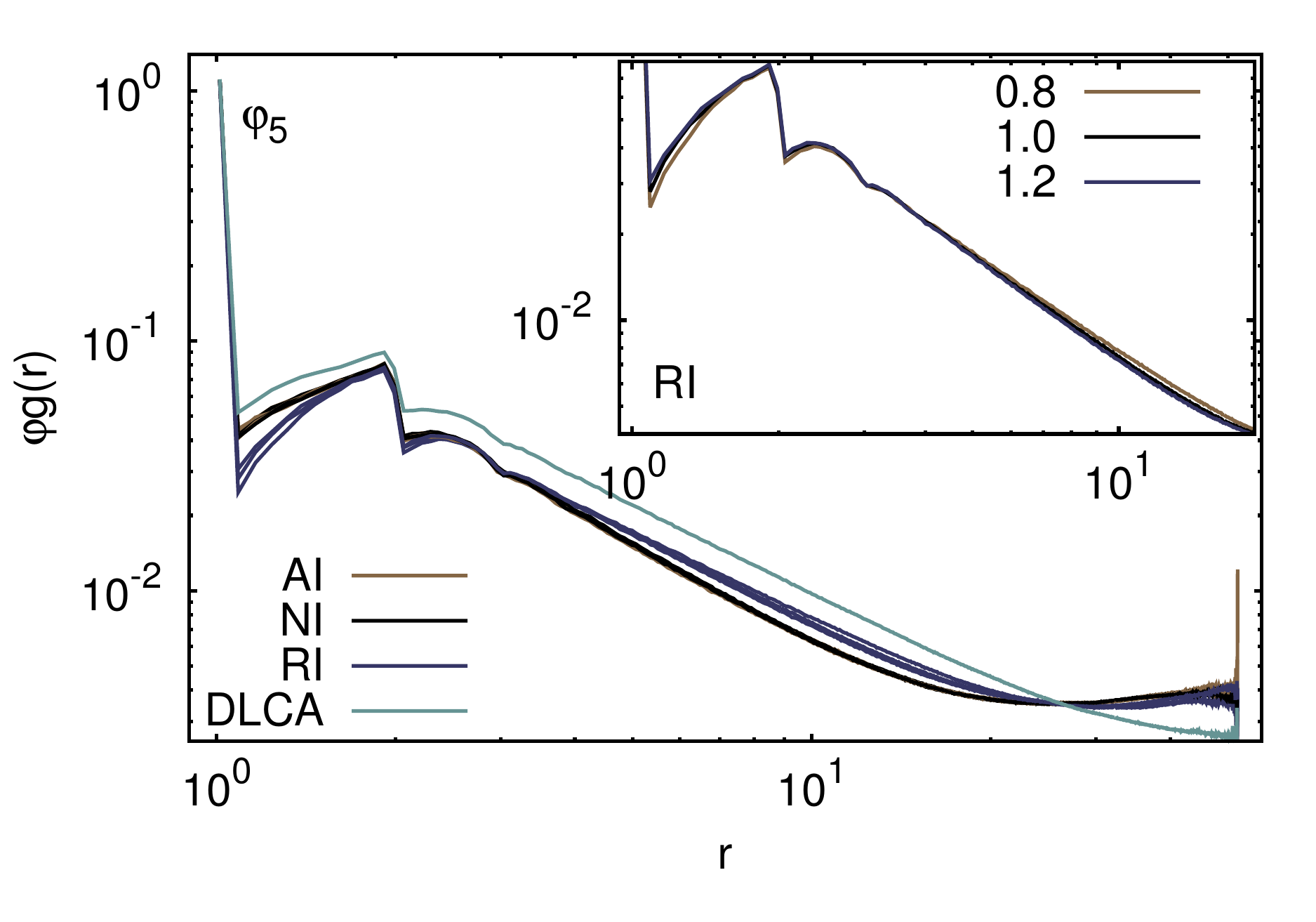}
\caption{\label{fig:gofronedensities} Pair distribution function $\varphi g(r)$ for the final aggregates formed by attractive (AI), repulsive (RI), and non-interacting (NI) particles at $T=0.8$, $1.0$, and $1.2$ (indistinguishable for attractive and non-interacting particles). The data for DLCA without rotational diffusion are presented for $T=1.0$ only. The inset illustrates the temperature dependence of $\varphi g(r)$ for repulsive particles. Note that the data are presented for a representative initial particle volume fraction $\varphi_5$ and the independence of the pair distribution function of the initial density for respective interparticle interactions is demonstrated in the supplementary material.} 
\end{center}
\end{figure}

The distribution of the particles inside the aggregates can further be analyzed in terms of the pair distribution function that characterizes the local density fluctuations by providing the average number of particles (multiplied here with the volume of a particle) in an element of volume at a certain distance from any particle. Essentially, the pair distribution function $\varphi g(r)$ is identical to the commonly used radial distribution function $g(r)$ but is, in this form, partially independent of the volume fraction of particles initially present in the system. Data presented in the supplementary material demonstrate that, while the radial distribution function depends on the initial density of the system, the values of the pair distribution function are locally identical for all volume fractions $\varphi_i$ considered. When the number of aggregating particles becomes sufficient to form percolating networks, the pair distribution function starts to vary with the volume fraction of particles initially present in the system. 
In Fig.~\ref{fig:gofronedensities}, we plotted the pair distribution functions for a representative volume fraction and various interparticle interactions at different temperatures together with the results obtained for the aggregates formed by non-interacting particles without rotational diffusion. Evidently, the absence of rotational diffusion leads to the formation of more compact structures, which can also be inferred from the respective radius of gyration. It is, however, remarkable that repulsive particles aggregate into structures which are locally less dense but on the large scale smaller than those formed by non-interacting and attractive particles. In accordance with the previous conclusions, we observe that the latter two aggregate into structures which are indistinguishable and independent of the temperature. The differences in the pair distribution functions with and without repulsive interactions illustrate an increase of the interpenetration depth of the aggregates formed by the repulsive particles. In the framework of RLCA, the compactness of the structures in comparison to DLCA is attributed to the possibility of two clusters to diffuse further into each other without forming a bond. In our case, the probability for two aggregates to combine is determined by the thermal energy of the particles and clusters which, in order to form an irreversible bond, should be sufficient to overcome the interparticle repulsion. The inset of Fig.~\ref{fig:gofronedensities} demonstrates that a temperature decrease indeed results in an increase of the number of particles found at intermediate distance range from any particle selected. In contrast, the number of particles found in the vicinity of each other decreases with temperature decrease.    

\begin{figure}[tb]
\begin{center}
\includegraphics[clip=,width=0.99\columnwidth]{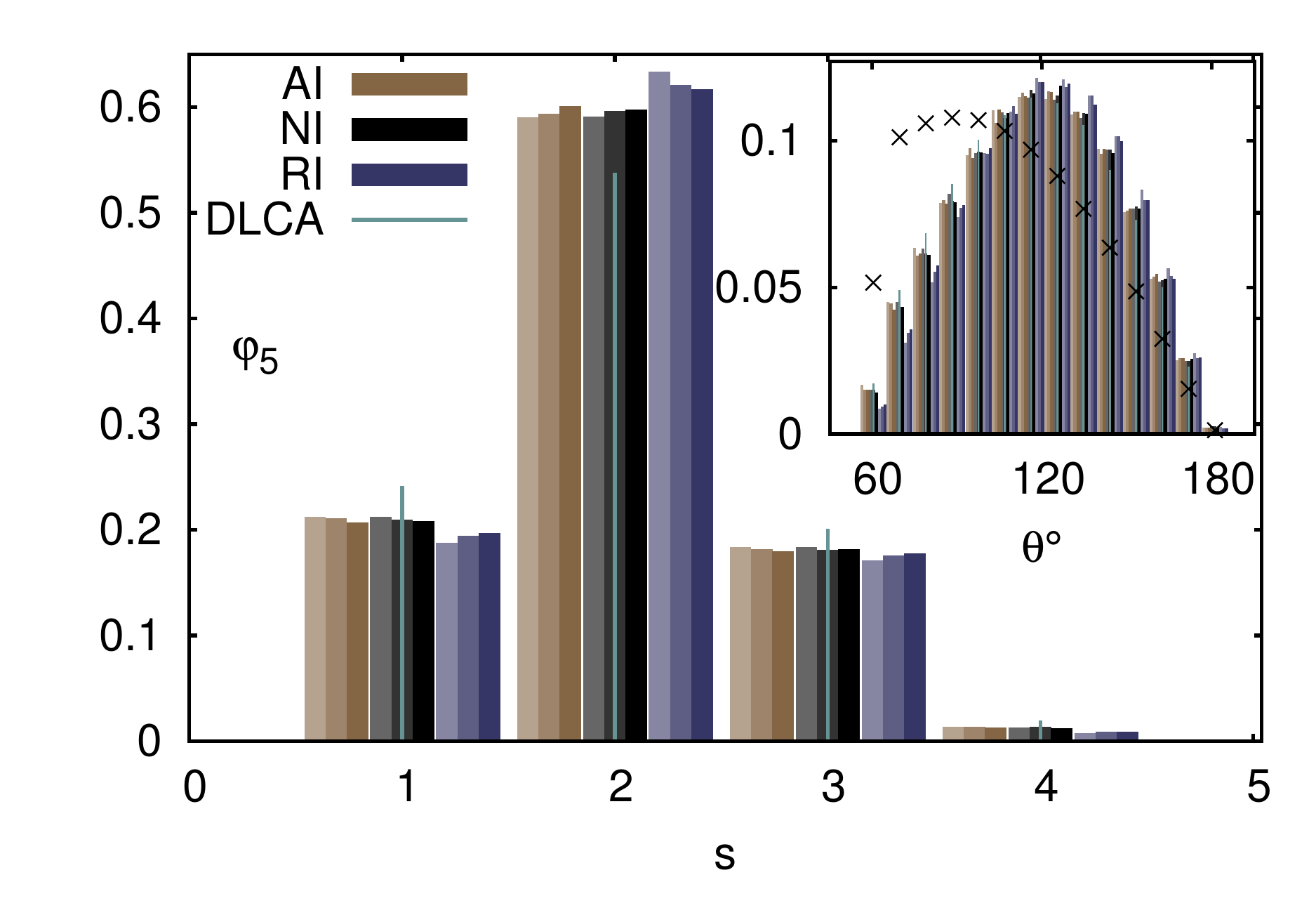}
\caption{\label{fig:nnonedensities} Distribution of the numbers of next neighbors of each particle in the final aggregates formed by attractive (AI), repulsive (RI), and non-interacting (NI) particles at $T=0.8$, $1.0$, and $T=1.2$ (the shade of the respective boxes increases with temperature). The data for DLCA without rotational diffusion are presented for $T=1.0$ only. Inset: Distribution of angles between the vectors connecting a chain particle ($s=2$) with its next neighbors. Crosses stand for the random distribution truncated at $\theta=60^{\circ}$. Note that the data are presented for a representative initial volume fraction $\varphi_5$ and the independence of the aggregate structure of the initial density for respective interparticle interactions is demonstrated in the supplementary material.} 
\end{center}
\end{figure}

A closer look at the surroundings of each particle in terms of the number of next neighbors, presented in Fig.~\ref{fig:nnonedensities}, reveals that the repulsive particles, in contrast to their attractive and non-interacting equivalents, tend to arrange in chains with less loose ends ($s=1$) and junctions ($s>2$). Figure~\ref{fig:nnonedensities} further confirms that the local structures formed by the attractive and non-interacting particles are similar, as already demonstrated by the respective pair distribution functions. The non-rotating clusters aggregated via the standard DLCA are characterized by significantly more intercepting chains and loose ends. The distribution of angles between the vectors connecting a particle in a chain ($s=2$) to its neighbors indicates that repulsive particles aggregate into chains which are more linear than those formed by the attractive and non-interacting particles. The temperature effects are also different: an increase of the temperature is associated either with more loose ends and junctions when the interparticle interactions are repulsive or with more attractive and non-interacting particles arranged into chains. A comparison with the random distribution of the next neighbor vectors, restricted to the angles producing no overlap between the neighbors of a particle, reveals that all chains are more elongated than in the case of randomly distributed contacts. 

\subsection{Aggregation times} 
\begin{figure}[tb]
\begin{center}
\includegraphics[clip=,width=0.99\columnwidth]{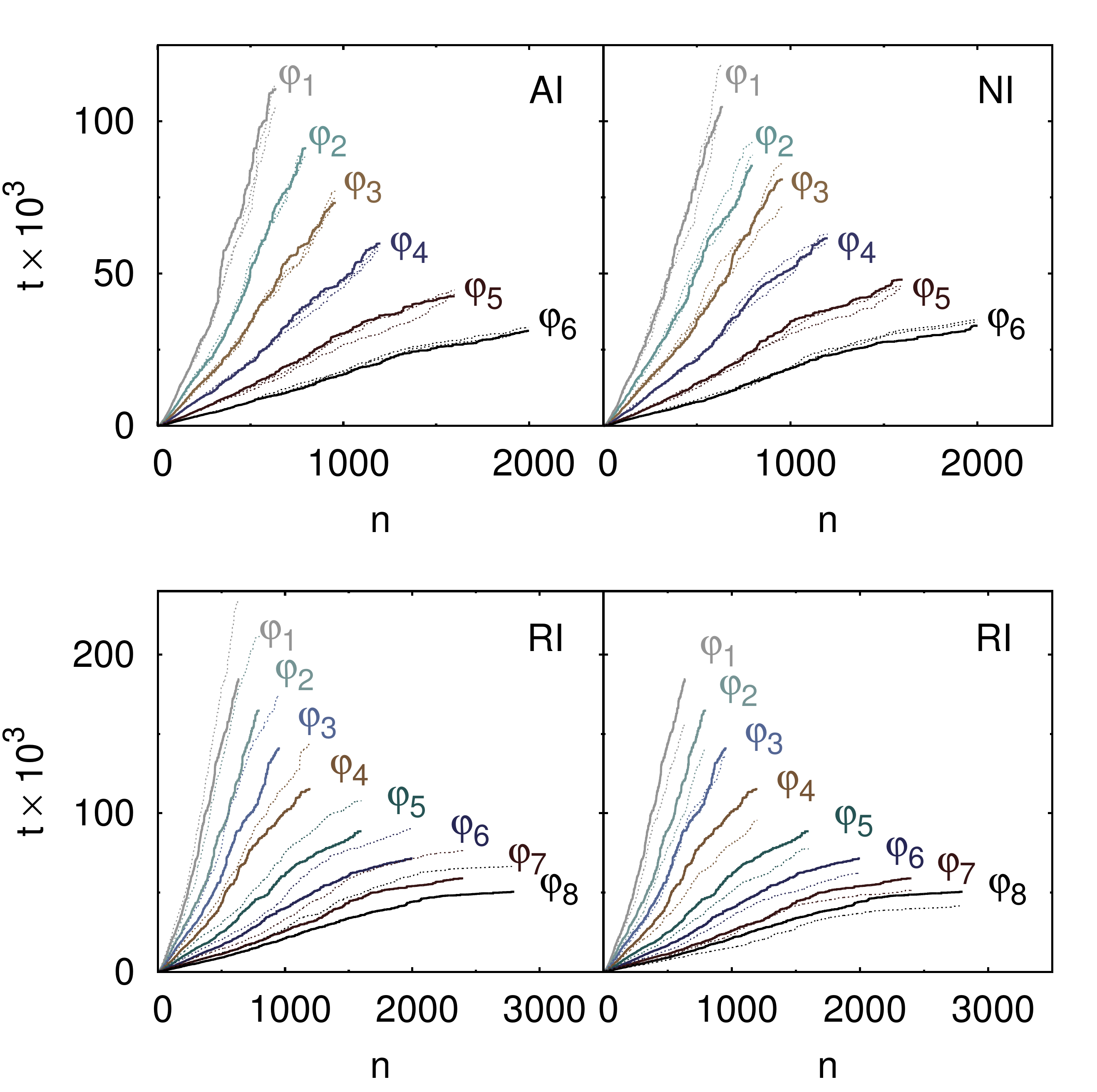}
\caption{\label{fig:times} Averaged time needed to form aggregates of a certain size for various initial volume fraction (as labeled). Note the different scales on the axis of ordinates. Top: Aggregates formed by attractive (left) and non-interacting (right) particles at $T=1.0$ (solid lines). Dotted lines stand for $T=0.8$ and $1.2$ indicating that the rescaled aggregation times of attractive and non-interacting particles are independent of the temperature. Bottom: Aggregation times of clusters formed by repulsive particles at $T=1.0$ (solid lines). Dotted lines stand for $T=0.8$ (left) and $1.2$ (right) demonstrating the variation of the aggregation times with temperature characteristic for the RLCA. } 
\end{center}
\end{figure}

Last but not least, we present, in Fig.~\ref{fig:times}, the aggregation times of clusters formed by the attractive, non-interacting, and repulsive particles. Attractive and non-interacting particles aggregate on the same timescale, set by the volume fraction of particles initially present in the system, which indicates that the attraction is too short-ranged to influence the diffusion of the aggregating clusters on the large scale. The rescaled aggregation times, which incorporate the temperature-induced variations of the diffusion process, are further independent of the temperature. In contrast, the times needed to form clusters from repulsive particles are considerably longer. An increase of the aggregation times with decreasing temperature, observed by the formation of clusters of repulsive particles, is characteristic for the RLCA. In this case, particles and clusters of particles diffuse to each other against a repulsive potential and combine into an aggregate only if their thermal energy is sufficient to overcome the strength of the interparticle repulsion. Hence, the length of the time periods between the formation of irreversible bonds is, for repulsive particles, determined by the temperature. 
      
\section{Conclusions} \label{sec:sum}

We have studied the process of irreversible aggregation of attractive, repulsive, and non-interacting nanoparticles. The aggregation of non-interacting particles represents an example of the DLCA of rotating clusters, for which we have shown recently \cite{jungblut:2019} that the structure of the aggregates depends on the ratio of rotational and translational diffusion coefficients. The formation of clusters by repulsive particles provides an instance of the RLCA, again including rotational diffusion. The findings presented in this article allow a new interpretation of the earlier experimental results \cite{weitz:1984a, weitz:1984b}, which supported the standard scenario of DLCA without rotational diffusion. We have shown, namely, that the fractal dimension of gold colloids' clusters, measured in experiments, can be obtained either by neglecting the effects due to the rotational diffusion of non-interacting particles or by considering aggregation of weakly repulsive particles and rotating clusters. Combining these results with the observation that the clusters obtained experimentally diffused both translationally and rotationally, we conclude that the porosity of the fractal structures of irreversibly aggregating particles can be increased above the standard DLCA limit by further screening of the interparticle interactions. 
 
What is more, our results on the aggregation of attractive particles suggest that the same fractal structures can be obtained for clusters formed by the attractive and non-interacting particles. These findings contradict the conclusions of a number of previous studies \cite{kim:2000a,puertas:2001,kim:2003}, which associated the decrease of the fractal dimension beyond the conventional DLCA with interparticle attraction. In this works, however, less compact structures were realized by heteroaggregation of $1:1$ mixtures of positively and negatively charged particles. Evidently, the high porosity of the structures formed in this case should be attributed not to the interparticle attraction, which does not change the structures of the aggregates, but to the repulsion between a particle and its second-shell neighbors. The interplay between short-ranged attraction and long-ranged repulsion has been previously studied in the context of the formation of equilibrium gels \cite{sciortino:2005gels,valadezPerez:2013a,cruz:2016} but not yet related to the process of irreversible heteroaggregation. 

\section*{Supplementary Material} 
See supplementary material for the demonstration that the local structure of the aggregates is independent of the volume fraction of particles initially present in the system.  

\acknowledgments

We gratefully acknowledge financial support from the European Research Council (ERC-2013-AdG AEROCAT). Furthermore, JOJ acknowledges financial support by the Deutsche Forschungsgemeinschaft (DFG) for M-era.NET project ICENAP. The computations were performed on an HPC system at the Center for Information Services and High Performance Computing (ZIH) at TU Dresden within the project QDSIM.

\balancecolsandclearpage
\begin{center}
\textbf{\Large Supplementary Material: Diffusion- and reaction-limited cluster aggregation revisited}
\end{center}
 
\setcounter{section}{0}
\setcounter{subsection}{0}
\setcounter{equation}{0}
\setcounter{figure}{0}
\setcounter{table}{0}
\setcounter{page}{1}
\makeatletter
\renewcommand{\theequation}{S\arabic{equation}}
\renewcommand{\thefigure}{S\arabic{figure}}
\renewcommand{\bibnumfmt}[1]{[S#1]}
\renewcommand{\citenumfont}[1]{S#1}
 
\section*{Local structure of final aggregates}
\begin{figure}[b]
\begin{center}
\includegraphics[clip=,width=0.8\columnwidth]{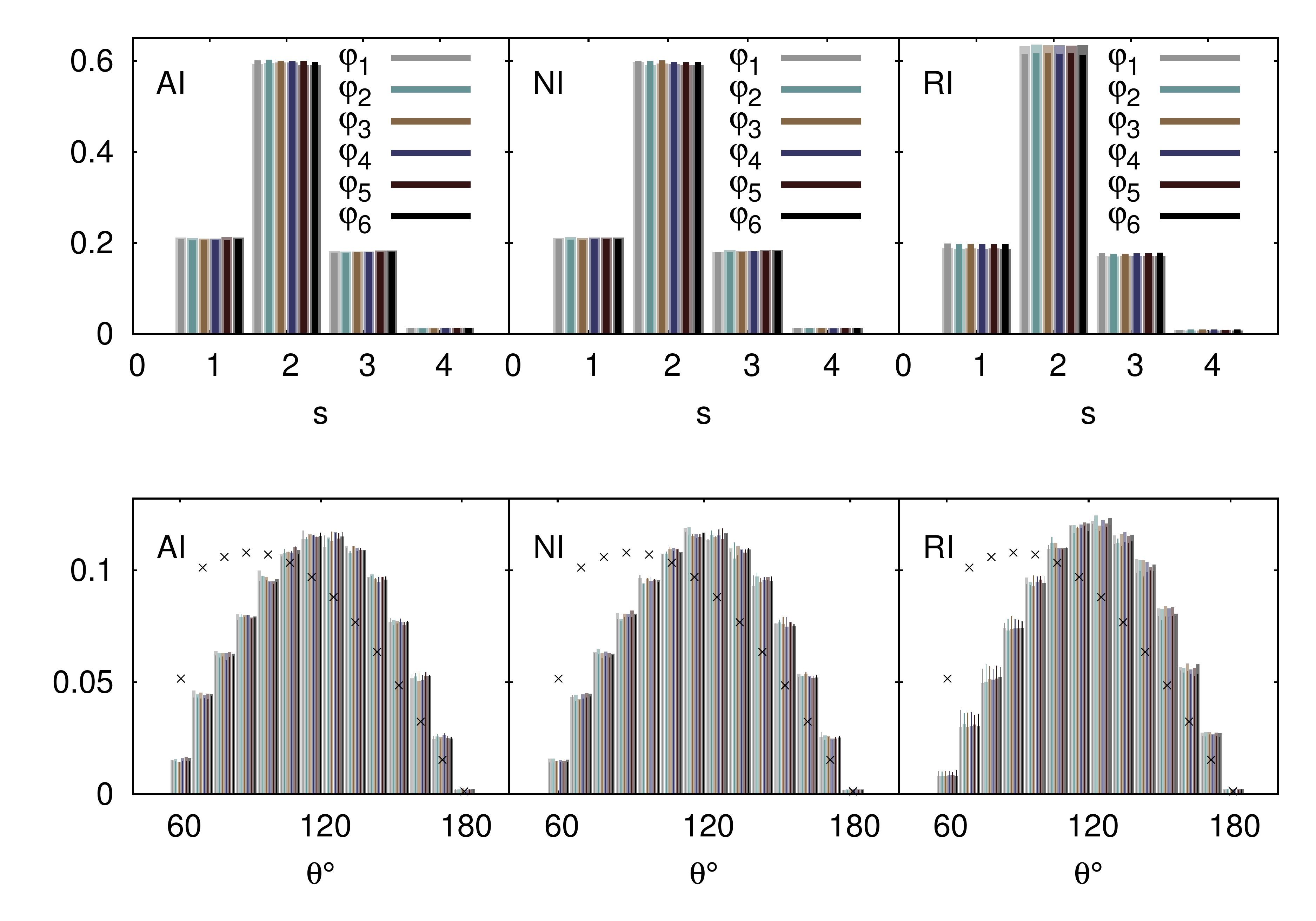} 
\caption{\label{fig:nnalldensities} Distribution of the numbers of next neighbors of each particle (top row) and of the angles between the vectors connecting a particle ($s=2$) with its next neighbors (bottom row) in the final aggregates formed by attractive (left), repulsive (right), and non-interacting (middle) particles at $T=0.8$ (shaded boxes) and $T=1.2$ (solid impulses). Crosses stand for the random distribution truncated at $\theta=60^{\circ}$. } 
\end{center}
\end{figure}

\begin{figure}[tb]
\begin{center}
\includegraphics[clip=,width=0.8\columnwidth]{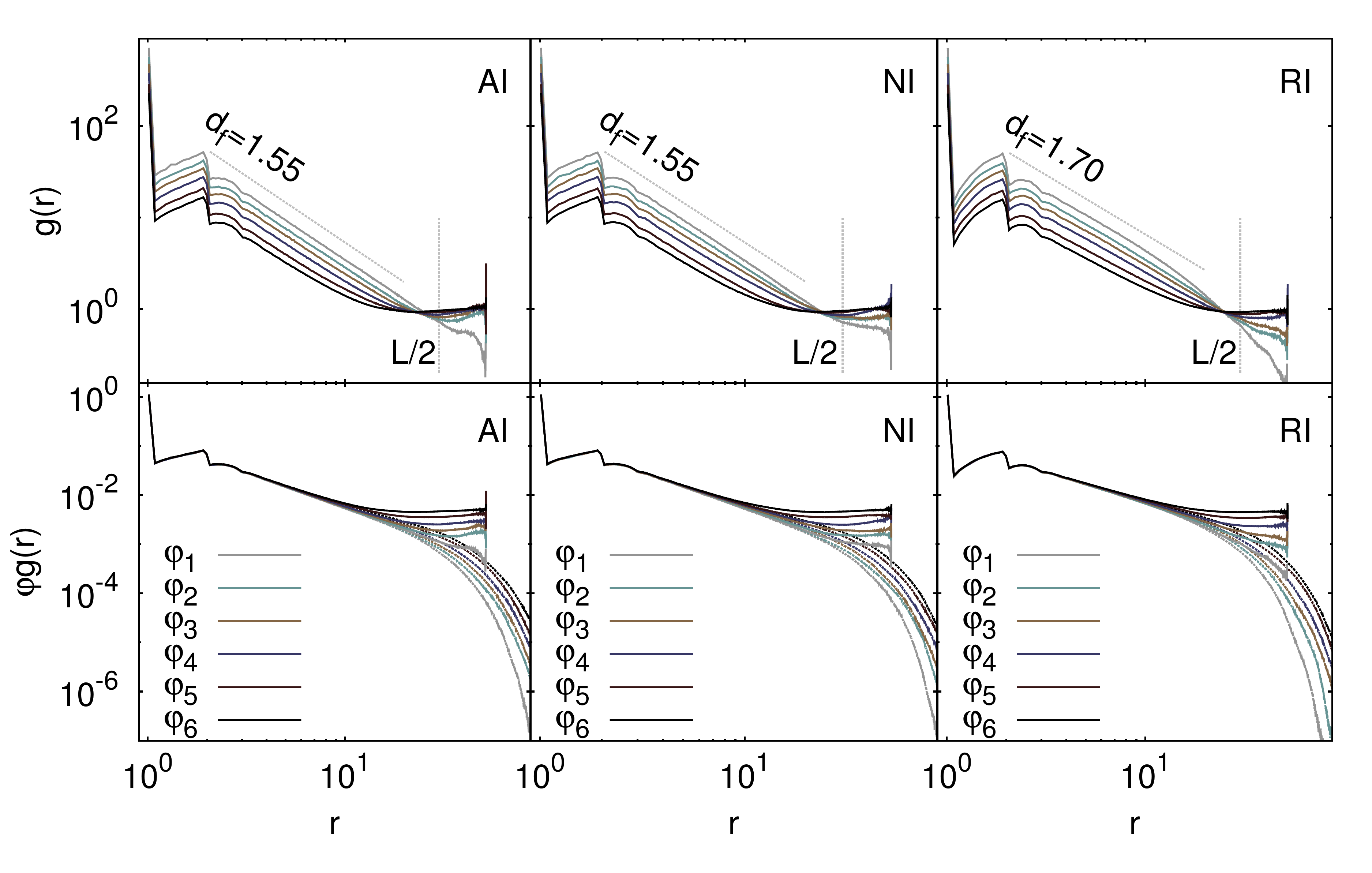} 
\caption{\label{fig:rofgalldensities1} Radial (top row) and pair (bottom row) distribution functions for the final aggregates formed by attractive (AI), repulsive (RI), and non-interacting (NI) particles at $T=0.8$. Pair distribution functions are computed with (solid lines) and without (broken lines) periodic boundary conditions. } 
\end{center}
\end{figure}

\begin{figure}[tb]
\begin{center}
\includegraphics[clip=,width=0.8\columnwidth]{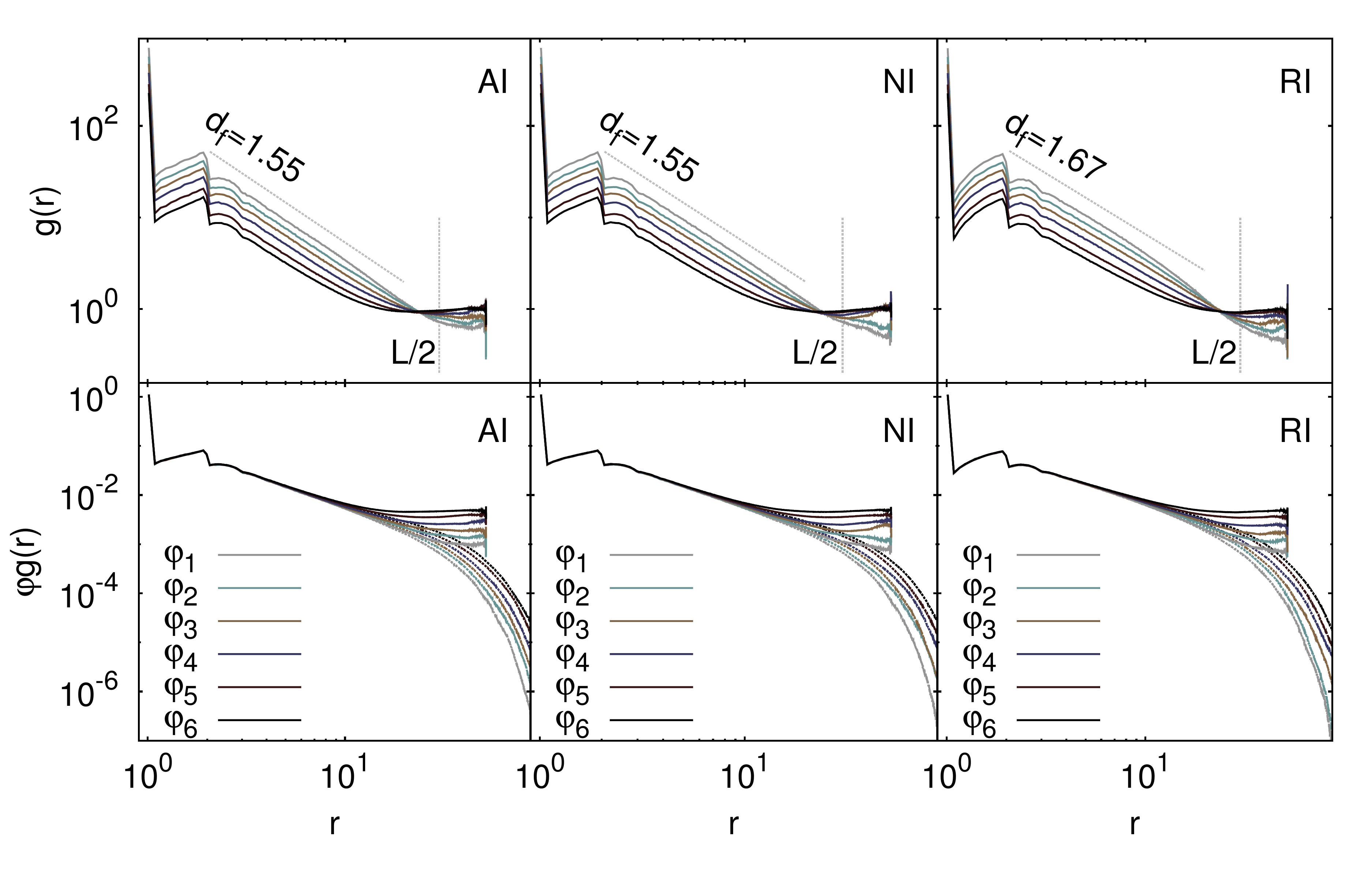}
\caption{\label{fig:rofgalldensities2} Same as Fig.~\ref{fig:rofgalldensities1} but for $T=1.0$.} 
\end{center}
\end{figure}

\begin{figure}[tb]
\begin{center}
\includegraphics[clip=,width=0.8\columnwidth]{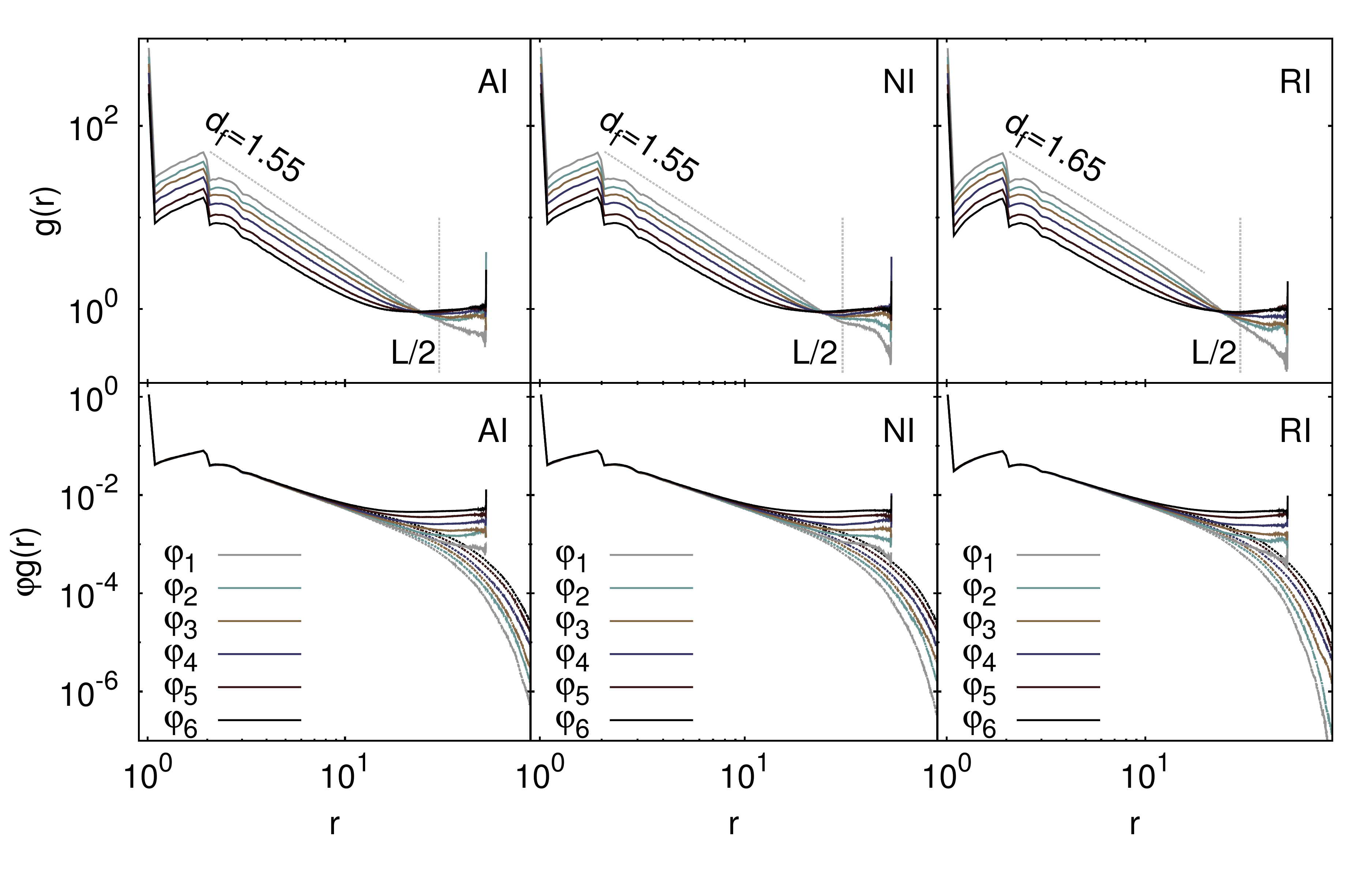} 
\caption{\label{fig:rofgalldensities3} Same as Fig.~\ref{fig:rofgalldensities1} but for $T=1.2$.} 
\end{center}
\end{figure}

The local structure of final aggregates does not depend on the volume fraction of the particles initially present in the system. In particular, Fig.~\ref{fig:nnalldensities} demonstrates that, for all interparticle interactions considered, the distribution of the next neighbors of a particle in a final aggregate does not change with the variation of the initial particle volume fraction.  
Aside from that, Figs.~\ref{fig:rofgalldensities1}-\ref{fig:rofgalldensities3} show that, while the radial distribution function $g(r)$ varies with the initial density, the pair distribution function $\varphi g(r)$ is locally independent of the particle concentration. Evidently, this independence is realizable only on the length scales on which the aggregates do not overlap. In computer simulations, the occurrence of such overlaps is indicated by the deviation between the pair distribution functions computed with and without employing periodic boundary conditions. Another indication for the onset of percolation is the saturation of the radial distribution function $g(r)$ to unity at length scales comparable to the half of the simulation box size. At intermediate length scales, however, the radial distribution function evolves according to  
\begin{equation}
g(r)\propto r^{d_f-d} 
\end{equation}
and can be used to determine the fractal dimension of the structure from the fit of the scaling function to the data in the scaling region. In this work, we used another method described in the main text but demonstrate, in Figs.~\ref{fig:rofgalldensities1}-\ref{fig:rofgalldensities3}, that the scaling regime is clearly visible.     


\end{document}